\documentclass[a4paper]{article}
\usepackage{multirow}
\usepackage{ISCSLP_v2}
\title{Speech Enhancement Based on Reducing the Detail Portion of Speech Spectrograms in Modulation Domain via Discrete Wavelet Transform}
\name{Shih-kuang Lee$^1$, Syu-Siang  Wang$^2$, Yu Tsao$^2$, Jeih-weih Hung$^1$}
\address{
  $^1$National Chi Nan University, Taiwan\\
  $^2$Academia Sinica, Taiwan}
\email{s105323501@mail1.ncnu.edu.tw, sypdbhee@gmail.com, yu.tsao@citi.sinica.edu.tw, jwhung@ncnu.edu.tw}

\begin{document}

\maketitle
\begin{abstract}
 In this paper, we propose a novel speech enhancement (SE) method by exploiting the discrete wavelet transform (DWT). This new method reduces the amount of fast time-varying portion, viz. the DWT-wise detail component, in the spectrogram of speech signals so as to highlight the speech-dominant component and achieves better speech quality.  A particularity of this new method is that it  is completely unsupervised and requires no prior information about the clean speech and noise in the processed utterance. The presented DWT-based SE method with various scaling factors for the detail part is evaluated with a subset of Aurora-2 database, and the PESQ metric is used to indicate the quality of processed speech signals. The preliminary results show that the processed speech signals reveal a higher PESQ score in comparison with the original counterparts. Furthermore, we show that this method can still enhance the signal by totally discarding the detail part (setting the respective scaling factor to zero), revealing that the spectrogram can be down-sampled and thus compressed without the cost of lowered quality.  In addition, we integrate this new method with conventional speech enhancement algorithms, including spectral subtraction, Wiener filtering, and spectral MMSE estimation, and show that the resulting integration behaves better than the respective component method. As a result, this new method is quite effective in improving the speech quality and well additive to the other SE methods.

\end{abstract}

\section{Introduction}

In various speech-related applications, an input speech signal often suffers from environmental noise and requires further processing with a speech enhancement method to improve the associated quality before being of use.    By and large, speech enhancement methods can be divided into two classes: unsupervised and supervised. Unsupervised methods, such as spectral subtraction (SS) \cite{SS1,SS2,SS}, Wiener filtering \cite{WF1,WF}, short-time spectral amplitude (STSA) estimation \cite{STSA} and short-time log-spectral amplitude estimation (logSTSA) \cite{logSTSA}. By contrast, supervised speech enhancement methods use a training set to learn distinct models for clean speech and noise signals, which examples include codebook-based approaches \cite{SE_code} and hidden Markov model (HMM) based methods \cite{SE_HMM}. 

Conventional speech enhancement methods usually process a noisy utterance in a frame-wise manner, viz. to enhance each short-time period of the utterance nearly independently.  However, recent researches show that considering the inter-frame variation over a relatively long span of time can contribute to superior performance in speech enhancement. Some well-known methods along this direction include modulation domain spectral subtraction \cite{MDSS} , modulation-domain Wiener filtering and Kalman filtering \cite{MDWF, MDKF}. In addition, compared with the conventional Fourier transform in which only the frequency parts are considered, the discrete wavelet transform (DWT) \cite{DWT} takes care of both the time and frequency aspects of a signal and is becoming popular in speech analysis. For example, the well-known wavelet thresholding denoising (WTD) \cite{WTD} uses the wavelet transform to split the time-domain signal into sub-bands and then performs thresholding.  A recent research \cite{SSW} applies the DWT to the plain speech feature time series and simply keeps the obtained approximation portion, which achieves data compression and noise robustness in recognition simultaneously.

Partially inspired by the aforementioned ideas, in this study we propose to employ the discrete wavelet transform to analyze the spectrogram of a noisy utterance along the temporal axis, and then devalue the resulting detail portion with a hope to reduce noise effect in order to promote speech quality. In spite of the simplicity of its implementation, the preliminary evaluation results indicate that the proposed method can provide input signals with better perceptual quality. It is also shown that this new method can be paired with several well-known speech enhancement methods to achieve even better performance.

The remainder of this study is organized as follows: Section 2 gives the detailed procedure of the newly proposed method together with the associated discussions and a preliminary test using an example utterance. The experimental setup is described in Section 3, and Section 4 includes the detailed experimental results and the corresponding analyses. Finally, a brief concluding remark including the future avenues is stated in Section 5.
\begin{figure*}
\centering
  \includegraphics[width=0.9\textwidth]{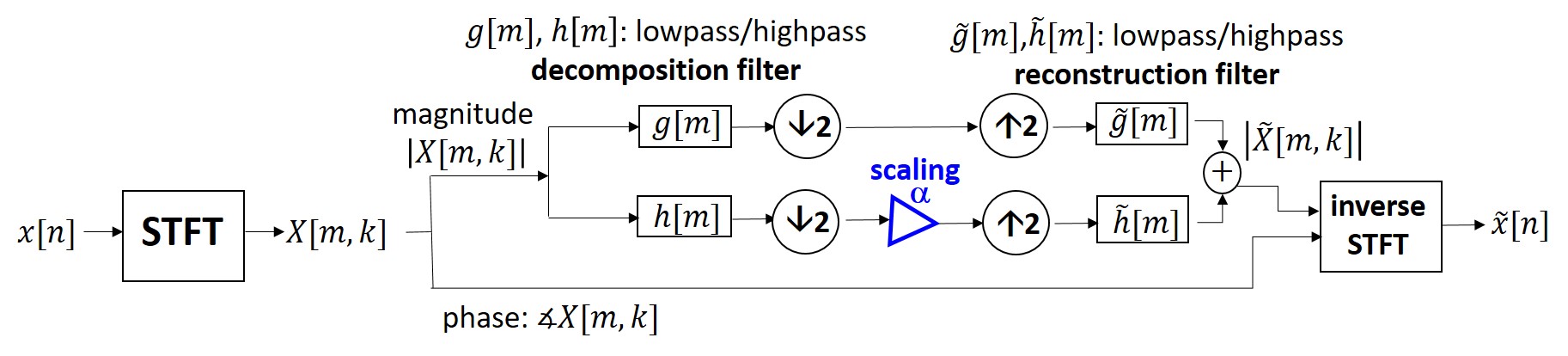}
  \caption{The flowchart of ModWD.}
    \label{fig:ModWD}
\end{figure*}
\section{Proposed method}
\begin{figure}[t]
  \centering
  \includegraphics[width=\columnwidth]{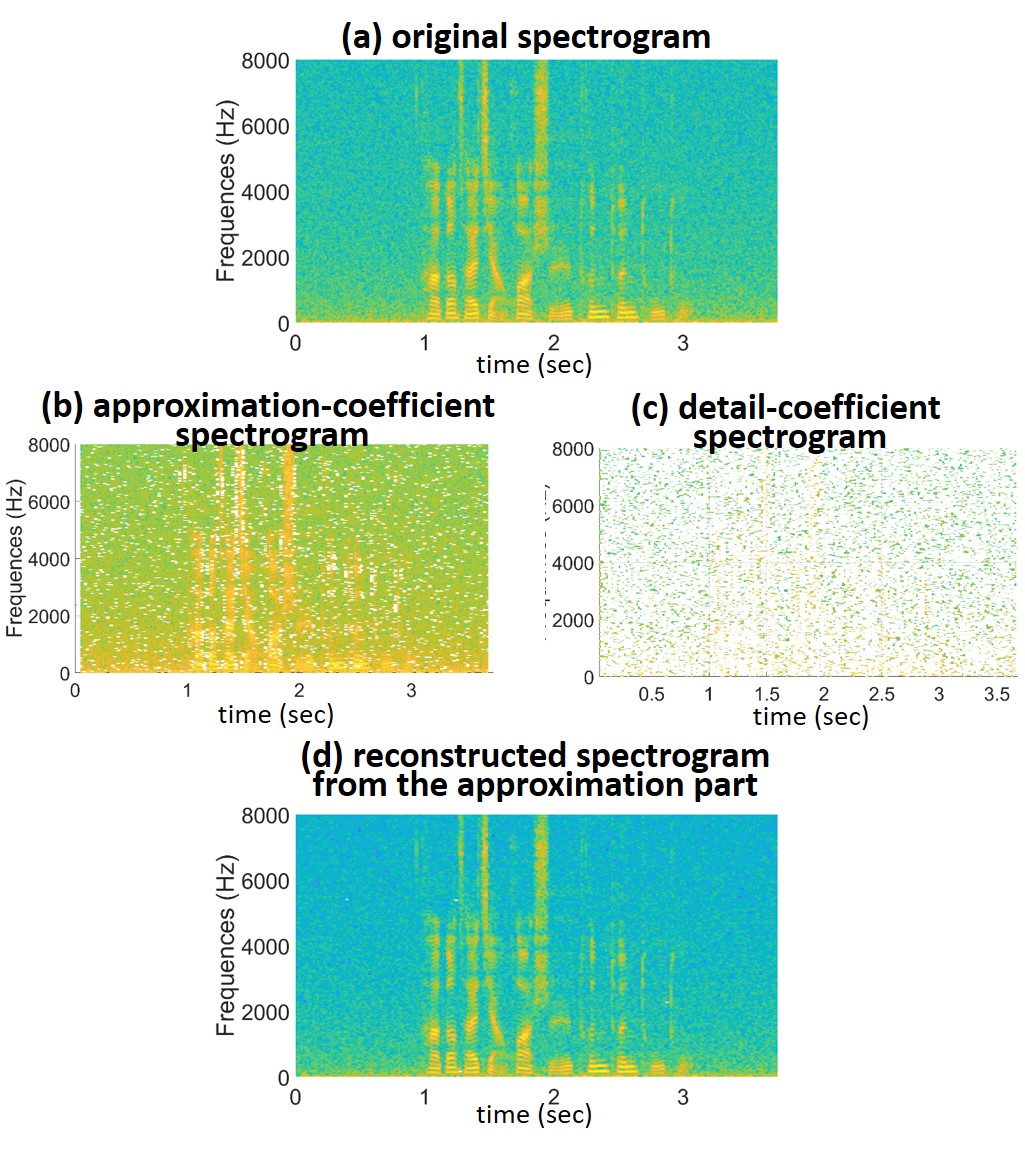}
  \caption{The different forms of spectrograms of an utterance in the process of ModWD with $\alpha=0$: (a) original spectrogram (b) approximation-coefficient spectrogram (c) detail-coefficient spectrogram (d) reconstructed spectrogram }
  \label{fig:spectrogram}
\end{figure}

Here, we present a novel speech enhancement method that employs the DWT to process the spectrogram of a speech signal. This method has a flowchart shown in Figure \ref{fig:ModWD} and consists of the following four steps:
\begin{enumerate}
\item[Step 1:] Create the spectrogram $\{X[m,k], 0\leq  m \leq L-1, 0\leq  k \leq K-1\}$ for a given time-domain signal $x[n]$, where $m$ and $k$ are respectively the indices of frame and acoustic frequency, and $L$ and $K$ are the total numbers of frames and acoustic frequency points, respectively.
\item[Step 2:] Use a one-level DWT to decompose the magnitude spectral sequence $\{|X[m,k]|, 0\leq  m \leq L-1\}$ with respect to any specific acoustic frequency index $k$. That is, we apply the DWT along with the horizontal axis of the spectrogram. The output of the one-level DWT consists of the approximation part $\{X_a [m,k]\}$ and the detail part $\{X_d[m,k]\}$, both of which have approximately half the length of the input $\{X [m,k]\}$ due to the factor-2 down-sampling.
\item[Step 3:] Reduce the detail part $\{X_d[m,k]\}$ by multiplying a factor $\alpha$ less than $1$ while keeping the approximation part $\{X_a [m,k]\}$ unchanged for the subsequent one-level inverse DWT (IDWT). That is, we feed the original approximation part and scaled detail part into the IDWT to reconstruct the magnitude spectral sequence $\{|\hat{X}[m,k]|\}$ associated with the acoustic frequency index $k$.
\item[Step 4:] The new time-domain signal $\tilde{x}[n]$ is created by applying the inverse STFT to the updated spectrogram, which consists of the new magnitude spectrogram$\{|\hat{X}[m,k]|\}$ and the original phase spectrogram $\{\angle X[m,k]\}$.
\end{enumerate}
The main concepts and characteristics of this new method are as follows:
\begin{enumerate}
	\item The DWT applied to the temporal sequence of magnitude spectrum is analogous to separating it into two sets of different modulation frequencies, each of which corresponds to a distinct rate of change in the temporal axis. Reducing the detail portion of the DWT output amounts to emphasizing the low modulation-frequency part of the original magnitude spectral sequence, and thus this new method is termed by the modulation-domain wavelet denoising, with a short-hand notation ModWD.
	\item According to \cite{mod_freq1}, for a speech signal the linguistic information is mainly located at the modulation frequencies within the range $1-16$ Hz, with a dominant modulation frequency of $4$ Hz. A plenty of research reports \cite{mod_freq2,mod_freq3, mod_freq4, mod_freq5, mod_freq6} also show that highlighting the relatively low modulation frequency components in the speech feature time series benefits the speech quality and recognition accuracy significantly in an adverse environment. As a result, we believe that alleviating the detail portion while preserving the approximation portion of the magnitude spectral sequence can further enhance the speech component and reduce noise effect.
	\item In addition to sub-band separation, another operation of the DWT is a factor-2 down-sampling. The newly proposed ModWD simply keeps the factor-2 down-sampled low-pass filtered sequence (the approximation portion) for the subsequent processing. When applying ModWD in a client-server transmission system, the operations of creating a spectrogram, one-level DWT and discarding the detail portion can be implemented on the client side, and the data being transmitted to the server side for further IDWT and reconstructing a time-domain signal is just half the size of the original spectrogram. Therefore, the presented ModWD is expected to improve the transmission efficiency without the cost of losing predominant speech information.

\end{enumerate}
Figures \ref{fig:spectrogram}(a)(b)(c)(d) depict the various types of magnitude spectrograms of an utterance during the process of ModWD. Comparing Figures \ref{fig:spectrogram}(a)(b)(c), the approximation-coefficient spectrogram is shown to contain much more correlation with the original spectrogram than the detail-coefficient spectrogram. Moreover, from Figures \ref{fig:spectrogram}(a)(d), the new spectrogram reconstructed from the approximation-coefficient spectrogram alone is quite close to the original spectrogram. Therefore, from these figures we provide a preliminary confirmation for the effectiveness of the proposed ModWD that highlights the speech-dominant component in speech purely depending on the DWT-wise approximation portion. 
\section{Experimental Setup}
This section presents the database, configurations of the speech enhancement systems, and the evaluation metric.

\subsection{Speech data preparation}
The experiments use the utterances included in the Aurora-2 database\cite{AURORA}, which contains connected English digit utterances generated by both female and male speakers at a sampling rate of 8 kHz. Parts of these utterances are contaminated by various types of noise at different SNRs. In the experiments, 50 airport-noise corrupted utterances belonging to a single speaker were used to form the test set. The SNR levels of the noise-corrupted utterances were varied from 0 dB to 20 dB, with a step of 5 dB. 

\subsection{Speech enhancement setup}
Some information about the setup of ModWD used in this study is as follows: 
\begin{itemize}
\item Each utterance was split into overlapped frames. The frame duration and frame shift were set to 20 ms and 10 ms, respectively. A Hamming window was then applied to each frame signal.
\item The number of frequency bins for the short-time Fourier transform (STFT) was set to 256.
\item The biorthogonal 3.7 wavelet basis was used for the DWT and inverse DWT of ModWD.
\item The scaling factor $\alpha$ used in ModWD was set to 0, 0.25, 0.5 and 0.75. 
\end{itemize}

\subsection{Objective evaluation metric}
Perceptual estimation of speech quality (PESQ) \cite{PESQ} was used as the evaluation metric. PESQ indicates the quality difference between the enhanced and clean speech signals, and it is analogous to the mean opinion score, which is a subjective evaluation index. The PESQ score ranges from 0.5 to 4.5, and a high score indicates that the enhanced utterance is close to the clean utterance. 

\section{Experimental results and discussions}

We first evaluated the proposed ModWD in its capability of enhancing noisy utterances. Then ModWD was integrated with several well-known speech enhancement methods to see if further improvement of speech quality could be achieved. 

\subsection{ModWD with various settings of the scaling factor $\alpha$}

 Table \ref{ModWD} lists the PESQ results with respect to the tested utterances for the baseline and the counterparts processed by ModWD at different values of the scaling factor  $\alpha$ for the detail portion. (Notably the unprocessed baseline is identical to ModWD with $\alpha=1$). From this table, some observations can be made:
\begin{enumerate}
    \item For all tested utterances, the associated PESQ score always gets lower as the SNR becomes worse, indicating that PESQ is an appropriate metric to reflect the degree of noisy distortion in speech.
    \item Compared with the baseline results, the proposed ModWD with the scaling factor $\alpha$ less than $1$ can achieve higher PESQ results for almost all tested utterances, with the only exception being ModWD with $\alpha=0$ at 0 dB SNR. Therefore, the potential of reducing noise effect of the proposed ModWD is  clearly revealed. As an aside, the possible explanation for the degraded performance of ModWD with $\alpha=0$ at 0 dB SNR is that significant extra distortion is introduced to a noisy spectrogram by completely discarding the associated detail (high-pass) portion.          
\item Setting the factor $\alpha$ to be either $0.25$ or $0$ in ModWD can give the best possible performance for the cases of SNR greater than 0 dB, which agrees with the findings in the past research \cite{mod_freq1} that high modulation frequency components in speech contain less linguistic information and are vulnerable to noise. Substantially reducing or even discarding these components can benefit the speech quality by reducing noise without the expense of introducing significant speech distortion.
\item ModWD with $\alpha=0$ indicates that only the DWT-wise approximation portion is required to participate in the subsequent inverse DWT since the detail portion is totally zeroed out. In this case, ModWD is computationally efficient for implementation and can achieve higher transmission efficiency in a client-server architecture because only the factor-2 down-sampled approximation portion has to be sent. 

\end{enumerate}

\begin{table}[]
\centering
\caption{PESQ results for ModWD with different assignments of the scaling factor $\alpha$}
\label{ModWD}
\begin{tabular}{cccccccc}
\hline
\textbf{SNR} & 0 & 5 & 10 &  15 & 20 & Avg.\\
\hline
baseline & 1.300 &	1.768 &	2.060 &	2.391 &	2.780 &	2.060 \\
$\alpha=0.75 $ & 1.307 &	1.779 &	2.070 &	2.404 &	2.789 &	2.070 \\
$\alpha=0.50$ & 1.319 &	1.788 &	2.078 &	2.415 &	2.795 &	2.079 \\
$\alpha=0.25$ & 1.322 &	1.794 &	2.083 &	2.422 &	2.797 &	2.084 \\
$\alpha=0 $ &1.289  &	1.798 &	2.086 &	2.428 &	2.797 &	2.079 \\
\hline
\end{tabular}
\end{table}
%
%
\subsection{Cascading ModWD and other speech enhancement methods}
Next, we integrated ModWD with several well-known speech enhancement algorithms to see whether such integration can further improve the quality of the testing utterances compared with the individual component method. The algorithms to be integrated include multi-band spectral subtraction (SS) \cite{SS}, Wiener filtering (WF) \cite{WF}, short-time log-spectral amplitude estimation (STSA) \cite{STSA} and short-time log-spectral amplitude estimation (logSTSA) \cite{logSTSA}. Please note that interchanging the cascading order of any two algorithms discussed here will behave differently since they are non-linear operations. As for the notations used later,  both ``$A-B$'' and ``$B-A$'' are the cascade of methods $A$ and $B$, while ``$A-B$''  indicates performing method $A$ first and then method $B$, and ``$B-A$''  goes the other way around. 

\begin{figure}[t]
  \centering
  \includegraphics[width=0.9\columnwidth]{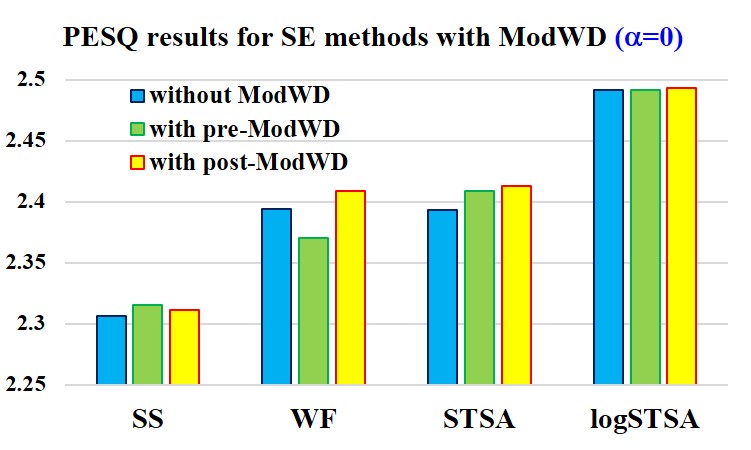}
  \caption{PESQ results averaged over five SNR cases for various SE methods with or without ModWD ($\alpha=0$)}
  \label{fig:PESQ1}
\end{figure}

\begin{figure}[t]
  \centering
  \includegraphics[width=0.9\columnwidth]{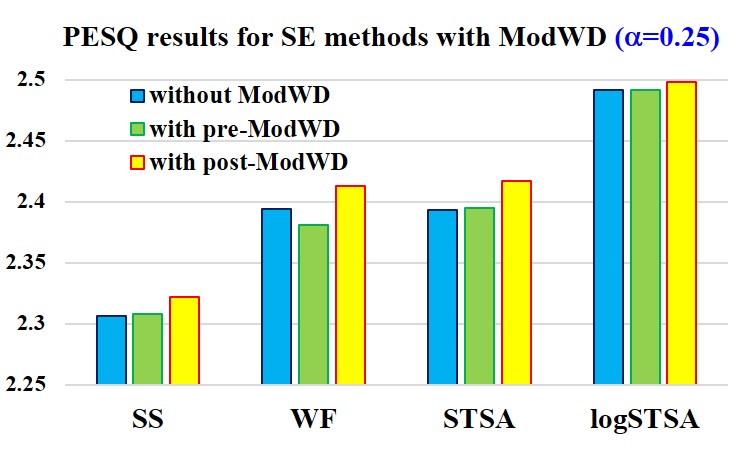}
  \caption{PESQ results averaged over five SNR cases for various SE methods with or without ModWD ($\alpha=0.25$)}
  \label{fig:PESQ2}
\end{figure}

The PESQ results for the integration of ModWD and any of SS, WF, STSA and  logSTSA are listed in Tables \ref{ModWD_SE1} and \ref{ModWD_SE2}. Here the scaling factor $\alpha$ of ModWD is set to either of $0$ and $0.25$. Figures \ref{fig:PESQ1} and \ref{fig:PESQ2} further summarize these PESQ values averaged over different SNR cases for ease of comparison.  From these tables and figures, we have the following findings:
\begin{enumerate}
\item In terms of the PESQ values averaged over the five SNR cases achieved by each individual method, logSTSA behaves the best, followed by WF, STSA, SS and ModWD in turn. It is not surprising that the improvement brought by ModWD is relatively insignificant because ModWD does not have an explicit noise estimation and reduction procedure as the other methods.

\item Cascading ModWD with any of the four methods gives rise to better results than the individual component method in almost all cases, revealing that ModWD is well additive to speech enhancement algorithms so as to further imporve the speech quality in adverse environments.

\item For any of the four SE methods discussed here, ModWD serves as a post-processing stage better than it is used for pre-processing. A possible underlying reason for this result is ModWD tends to undermine the noise estimation accuracy of the cascaded SE method afterwards.

\item ModWD with $\alpha=0.25$ outperforms ModWD with $\alpha=0$ when they are integrated with any other SE method. However, the PESQ performance difference is marginal and less than $0.01$ in most cases.

\end{enumerate}

\begin{table}[]
\small
\centering
\caption{PESQ results for various SE methods including ModWD with $\alpha=0$}
\label{ModWD_SE1}
\begin{tabular}{cccccccc}
\hline
\textbf{SNR} & 0 & 5 & 10 &  15 & 20 & Avg.\\
\hline
ModWD          & 1.289 & 	1.798 & 	2.086 & 	2.428 &  	2.797 &  	2.079  \\
SS      	   & 1.489 & 	2.025 & 	2.341 & 	2.668 & 	3.007 & 	2.306  \\
WF	           & 1.723 & 	2.126 & 	2.412 & 	2.726 & 	2.983 & 	2.394  \\
STSA           & 1.700 & 	2.141 & 	2.430 & 	2.724 & 	2.974 & 	2.393  \\
logSTSA       & 1.840 & 	2.234 & 	2.523 & 	2.802 & 	3.058 & 	2.491  \\
\hline
ModWD       & \multirow{2}{*}{1.479} & 	\multirow{2}{*}{2.029} & 	\multirow{2}{*}{2.366} & 	\multirow{2}{*}{2.688} & 	\multirow{2}{*}{3.014} & 	\multirow{2}{*}{2.315}  \\
-SS        &  & 	 & 	 & 	 & 	 & 	\\
\hline
SS       & \multirow{2}{*}{1.500} & 	\multirow{2}{*}{2.048} & 	\multirow{2}{*}{2.317} & 	\multirow{2}{*}{2.694} & 	\multirow{2}{*}{2.998} & 	\multirow{2}{*}{2.311}  \\
-ModWD        &  & 	 & 	 & 	 & 	 & 	\\
\hline
ModWD	   & \multirow{2}{*}{1.676} & 	\multirow{2}{*}{2.079} & 	\multirow{2}{*}{2.409} & 	\multirow{2}{*}{2.720} & 	\multirow{2}{*}{2.967} & 	\multirow{2}{*}{2.370}  \\
-WF        &  & 	 & 	 & 	 & 	 & 	\\
\hline
WF       & \multirow{2}{*}{1.772} & 	\multirow{2}{*}{2.133} & 	\multirow{2}{*}{2.441} & 	\multirow{2}{*}{2.737} & 	\multirow{2}{*}{2.963} & 	\multirow{2}{*}{2.409}  \\
-ModWD        &  & 	 & 	 & 	 & 	 & 	\\
\hline
ModWD     & \multirow{2}{*}{1.696} & 	\multirow{2}{*}{2.156} & 	\multirow{2}{*}{2.449} & 	\multirow{2}{*}{2.749} & 	\multirow{2}{*}{2.993} & 	\multirow{2}{*}{2.409}  \\
-STSA        &  & 	 & 	 & 	 & 	 & 	\\
\hline
STSA     & \multirow{2}{*}{1.735} & 	\multirow{2}{*}{2.169} & 	\multirow{2}{*}{2.445} & 	\multirow{2}{*}{2.741} & 	\multirow{2}{*}{2.977} & 	\multirow{2}{*}{2.413}  \\
-ModWD        &  & 	 & 	 & 	 & 	 & 	\\
\hline
ModWD&  \multirow{2}{*}{1.813} & 	 \multirow{2}{*}{2.227} & 	 \multirow{2}{*}{2.520} & 	 \multirow{2}{*}{2.819} & 	 \multirow{2}{*}{3.076} & 	 \multirow{2}{*}{2.491}  \\
-logSTSA        &  & 	 & 	 & 	 & 	 & 	\\
\hline
logSTSA &  \multirow{2}{*}{1.842} & 	 \multirow{2}{*}{2.240} & 	 \multirow{2}{*}{2.519} & 	 \multirow{2}{*}{2.810} & 	 \multirow{2}{*}{3.054} & 	 \multirow{2}{*}{2.493}  \\
-ModWD        &  & 	 & 	 & 	 & 	 & 	\\
\hline
\end{tabular}
\end{table}

\begin{table}[]
\centering
\caption{PESQ results for various SE methods including ModWD with $\alpha=0.25$}
\label{ModWD_SE2}
\begin{tabular}{cccccccc}
\hline
\textbf{SNR} & 0 & 5 & 10 &  15 & 20 & Avg.\\
\hline
ModWD          & 	1.322 &	1.794 &	2.083 &	2.422 &	2.797 &	2.084  \\
SS      	   & 	1.489 &	2.025 &	2.341 &	2.668 &	3.007 &	2.306  \\
WF	            & 	1.723 &	2.126 &	2.412 &	2.726 &	2.983 &	2.394  \\
STSA            & 	1.700 &	2.141 &	2.430 &	2.724 &	2.974 &	2.393  \\
logSTSA        & 	1.840 &	2.234 &	2.523 &	2.802 &	3.058 &	2.491  \\
\hline
ModWD        & 	\multirow{2}{*}{1.496} &	\multirow{2}{*}{2.025} &	\multirow{2}{*}{2.341} &	\multirow{2}{*}{2.668} &	\multirow{2}{*}{3.007} &	\multirow{2}{*}{2.308}  \\
-SS        &  & 	 & 	 & 	 & 	 & 	\\
\hline
SS        & 	\multirow{2}{*}{1.502} &	\multirow{2}{*}{2.047} &	\multirow{2}{*}{2.364} &	\multirow{2}{*}{2.691} &	\multirow{2}{*}{3.005} &	\multirow{2}{*}{2.322}  \\
-ModWD        &  & 	 & 	 & 	 & 	 & 	\\
\hline
ModWD	    & 	\multirow{2}{*}{1.697} &	\multirow{2}{*}{2.087} &	\multirow{2}{*}{2.414} &	\multirow{2}{*}{2.721} &	\multirow{2}{*}{2.988} &	\multirow{2}{*}{2.381}  \\
-WF        &  & 	 & 	 & 	 & 	 & 	\\
\hline
WF        & 	\multirow{2}{*}{1.768} &	\multirow{2}{*}{2.133} &	\multirow{2}{*}{2.445} &	\multirow{2}{*}{2.742} &	\multirow{2}{*}{2.975} &	\multirow{2}{*}{2.413}  \\
-ModWD        &  & 	 & 	 & 	 & 	 & 	\\
\hline

ModWD  	 & 	\multirow{2}{*}{1.700} &	\multirow{2}{*}{2.149} &	\multirow{2}{*}{2.429} &	\multirow{2}{*}{2.724} &	\multirow{2}{*}{2.974} &	\multirow{2}{*}{2.395}   \\
-STSA        &  & 	 & 	 & 	 & 	 & 	\\
\hline
STSA  & 	 \multirow{2}{*}{1.740} &	\multirow{2}{*}{2.176} &	\multirow{2}{*}{2.445} &	\multirow{2}{*}{2.742} &	\multirow{2}{*}{2.982} &	\multirow{2}{*}{2.417}  \\
-ModWD        &  & 	 & 	 & 	 & 	 & 	\\
\hline
ModWD & 	 \multirow{2}{*}{1.840} &	 \multirow{2}{*}{2.234} &	 \multirow{2}{*}{2.523} &	 \multirow{2}{*}{2.802} &	 \multirow{2}{*}{3.057} &	 \multirow{2}{*}{2.491}  \\
-logSTSA        &  & 	 & 	 & 	 & 	 & 	\\
\hline
logSTSA  & 	 \multirow{2}{*}{1.846} &	 \multirow{2}{*}{2.246} &	 \multirow{2}{*}{2.522} &	 \multirow{2}{*}{2.813} &	 \multirow{2}{*}{3.062} &	 \multirow{2}{*}{2.498}  \\
-ModWD        &  & 	 & 	 & 	 & 	 & 	\\
\hline
\end{tabular}
\end{table}

\section{Conclusions}

This study presents a DWT-based speech enhancement approach that highlights the low modulation frequency components of the spectrogram so as to reduce the noise effect. In spite of no prior knowledge of the actual distortions adopted, the presented ModWD still improves the quality of utterances in unseen noise environments and is well additive to other speech enhancement methods. As to future work, we will adopt a multi-level DWT to achieve a high resolution in modulation frequency for the analyzed spectrogram and use  a validation set to learn the value of scaling factor $\alpha$ in order to further improve the effectiveness of ModWD.



\bibliographystyle{IEEEtran}

\bibliography{mybib}

\end{document}